\documentclass{aa}
\usepackage{graphicx}
\usepackage{epsfig}

\usepackage[varg]{txfonts}
\usepackage{url}
\usepackage[normalem]{ulem}
\usepackage{color}
\usepackage{hyperref}
\usepackage{color}

\newcommand{\msun}{M_{\sun}}
\newcommand{\rg}{R_{\rm S}}
\newcommand{\diff}{{\rm d}}

\newcommand{\unit}[1]{\mbox{\boldmath $\hat{#1}$}}

\newcommand{\beq}{\begin{eqnarray}}
\newcommand{\eeq}{\end{eqnarray}}
\newcommand{\be}{\begin{equation}}
\newcommand{\ee}{\end{equation}}
 
\newcommand{\lensing}{{\cal D}}
\newcommand{\Dop}{\delta}

\begin{document}

\title{Relativistic rotating vector model for X-ray millisecond pulsars} 
\titlerunning{Polarization from X-ray millisecond pulsars}

\author{Juri Poutanen \inst{1,2,3}}
\authorrunning{J. Poutanen}

\institute{
Department of Physics and Astronomy,  FI-20014 University of Turku, Finland
\\ 
\email{juri.poutanen@utu.fi}
\and
Space Research Institute of the Russian Academy of Sciences, Profsoyuznaya str. 84/32, 117997 Moscow, Russia 
\and
Nordita, KTH Royal Institute of Technology and Stockholm University, Roslagstullsbacken 23, 10691 Stockholm, Sweden}

\date{\today}

\abstract{The X-ray radiation produced on the surface of accreting  magnetised neutron stars is expected to be strongly polarised. 
A swing of the polarisation vector with the pulsar phase gives a direct measure of the source inclination and magnetic obliquity.
In the case of rapidly rotating millisecond pulsars, the relativistic motion of the emission region causes additional rotation of the polarisation plane.  
Here, we develop a relativistic rotating vector model, where we derive  analytical expression for the polarisation angle as a function of the pulsar phase  accounting for relativistic aberration and gravitational light bending in the Schwarzschild metric.
We show that in the case of fast pulsars the rotation of the polarisation plane can reach tens of degrees, strongly influencing the observed shape of the polarisation angle's phase dependence. 
The rotation angle grows nearly linearly with the spin rate but it is less sensitive to the neutron star radius. 
Overall, this angle is large even for large spots. 
Our results have implications with regard to the modelling of X-ray polarisation from accreting millisecond pulsars that are to be observed with the upcoming {Imaging X-ray Polarimeter Explorer} and the {enhanced X-ray Timing and Polarimetry} mission. 
The X-ray polarisation may improve  constraints on the neutron star mass and radius coming from the pulse profile modelling. 
}


   \maketitle
%

\section{Introduction}

Constraints on the equation of state of cold dense matter can be obtained using astrophysical measurements of neutron stars (NSs) \citep{LP07,OF16ARAA,Watts16RvMP,Watts19,Miller20}. 
The precise timing of radio pulsars  in binary systems \citep{Lorimer08} have constrained the maximum NS mass to be at least 2$M_\sun$ \citep{Demorest10,Antoniadis13,Fonseca16}, which rules out very soft equations of state.
The determination of NS radii would narrow down the possible equation of state and this has been the subject of numerous studies. 
For example, mass-radius relation can be obtained using spectral information on surface thermal emission during X-ray bursts \citep{Ozel09,SLB10,SPRW11,Steiner13,PNK14,Nattila16,Nattila17,Suleimanov20}, in quiescent states of NSs in binary systems \citep{Rutledge99,Heinke03,Heinke06,Webb07,Guillot13,Lattimer14,Bogdanov16,Steiner18}, as well as from central compact objects in supernova remnants \citep{Ho09,Klochkov15,Suleimanov17}. 
Additional constraints come from  tidal deformabilities measured using a gravitational wave signal from merging NSs \citep{Bauswein17,Annala18,Abbott18PhRvL,De18,Most18}. 

The parameters of a NS can also be measured from the X-ray pulse profiles of accreting  \citep{PG03,LMC08,Leahy09,ML11,SNP18}, nuclear \citep{BS05,lomiller13,ML15,Stevens16}, and radio millisecond pulsars \citep{Bogdanov08,Bogdanov13,Bogdanov19L26,MLD_nicer19,RWB_nicer19}.   
In particular, the  pulse profiles produced by a hotspot on the NS surface are strongly affected by both special relativistic (SR) and general relativistic (GR)  effects,  such as  the Doppler effect, light bending, and time delays and, therefore, they carry information about the NS radius and mass  \citep{PG03,PB06,CMLC07,MLCB07,NP18,Suleimanov20}. 
The magnetic field in these NSs is rather weak and does not affect the radiative transport, allowing us to theoretically predict the surface emission pattern and spectrum, as well as  the pulse profile, reliably. 
However, similar light curves can be produced with a very different set of parameters. For example, there is a degeneracy when exchanging the observer inclination and the magnetic obliquity \citep{VP04}.   
A possible way to distinguish between the models and to improve the constraints on the masses and radii is to observe variations of the polarisation degree (PD) and polarisation angle (PA) with the phase \citep{SNP18}.
These kinds of observations have turned out to be a valuable tool in determining the geometry of the radio pulsar emission region \citep[e.g.][]{Blaskiewicz91}. 
For X-ray pulsars, polarimetric observations will be possible in the near future with the  {Imaging X-ray Polarimeter Explorer (IXPE) } \citep{Weisskopf16SPIE}  and  the {enhanced X-ray Timing and Polarimetry (eXTP)}  mission \citep{Zhang16,Zhang19}. 
  
In this paper, we assume that the emission in the local frame corotating with the NS has azimuthal symmetry. 
This is the case for accreting millisecond pulsars, where most of radiation is produced at the magnetic poles, or for X-ray bursting NSs.
In both types of sources, the magnetic field plays no role in shaping the polarisation properties and the dominant direction of oscillations of the electric vector (aka the polarisation vector) has to lie in the plane containing the local normal or be perpendicular to it. 
Without losing the generality, we further consider the former case. 
For a slowly rotating star, the variations of the PA (i.e. the position angle of the electric vector) with the pulsar phase follow the projection of the hotspot normal on the sky, as described by the rotating vector model \citep{RC69}. 
However, when the source moves with a velocity that is a significant fraction of the speed of light, the preferred direction of the electric vector rotates.
A rather general expression for the rotation of the polarisation plane  was obtained by \citet{BL79} and \citet{LPB03} and discussed more recently by \citet{Lyutikov17PRD} and \citet{LK17MNRAS} with application to relativistic jets.  
The rotation of the polarisation plane of the radiation coming from the accretion discs around black holes was discussed by \citet{Pineault77}, \citet{Stark77}, \citet{Connors77}, and  \citet{Connors80}.
The latter paper also presented an analytical expression for the rotation angle, where only SR effects were considered. 

In the context of surface emission from a NS, the effect was considered  by \citet{Fer73,Fer76}, but no explicit analytical expression was found. 
A more recent work by \citet{Lyutikov16} also addressed this issue in the context of radio pulsars.   
None of the quoted papers have included the effects related to light bending  in the vicinity of a NS. 
Light bending was accounted for by \citet{Pavlov00} who considered, however, the case of slowly rotating, strongly magnetised pulsars. 
Numerical calculations of polarised radiation transfer from rapidly rotating NSs were recently presented by \citet{PMN18}, who also evaluated the effect of the deviation from the Schwarzschild metric. 

Here, we present relativistic rotating vector model that describes variations of the polarisation plane of emission produced by a hotspot on the surface of a rapidly rotating NS accounting for the relativistic aberration and gravitational light bending in the Schwarzschild metric.  
The results of the current study were already used by \citet{VP04}, but the derivation of the general expression was not presented there. 
Here, we consider only spherical stars, but taking on a generalisation to an oblate NS shape is straightforward and will be considered elsewhere \citep{loktev20}.
Another possible extension of our work is to pulsars with a strong magnetic field, where the polarisation vector may not lie in the plane containing the normal. 
The same formalism can also be used for the derivation of the rotation of the polarisation plane of the radiation from accretion discs around NSs and black holes. 

\section{Radiation from antipodal spots} 
\label{sec:flux_and_pol}

We consider a spherical NS of radius, $R$, and mass, $M$, and two identical antipodal hotspots at its surface displaced from the NS rotational axis. 
Due to stellar rotation, the visibility of the hotspots changes, resulting in variations of the observed flux, the PD, and the PA.
The observed flux and the PD at a given pulsar phase $\phi$ depends on the zenith angle $\alpha'$ between the momentum of emitted  photon and the local normal to the stellar surface in the spot comoving frame. 
We assume that they do not depend on the azimuthal angle at which the spot is seen in the comoving frame, which is a good approximation for low magnetic field pulsars (see \citealt{Pavlov00} for the high magnetic field case). 
The PD is invariant and does not change along the photon trajectory, thus, the observed PD is equal to that which is measured in the comoving frame at the emitted energy $E'$ and angle $\alpha'$.  
If we have a model to determine how Stokes parameters depend on energy and angles in the spot comoving frame, we can then  transform them to the observer frame. 
First, we make the Lorentz transformation to the non-rotating local frame, accounting for Doppler boosting and relativistic aberration. 
Then we follow photon trajectories to the observer at infinity in Schwarzschild space-time.
Deviations from the  Schwarzschild  metric due to the stellar rotation have a small effect both on the observed flux and polarisation \citep{BRR00,PMN18} and are neglected here. 
We also need to account for the time delays, which in the extreme cases of NS spinning at 600 Hz can reach 5--10 per cent of the pulsar period. 
 
For completeness, we first repeat the method which is used to compute the observed flux \citep[see also][]{PB06,SNP18,Bogdanov19L26,Suleimanov20} and then extend it to the description of the PD (Sect.~\ref{sec:obs_flux_PD}) and PA (in Sect.~\ref{sec:obs_PA}).

\begin{figure}
\begin{center}
\centerline{\epsfig{file=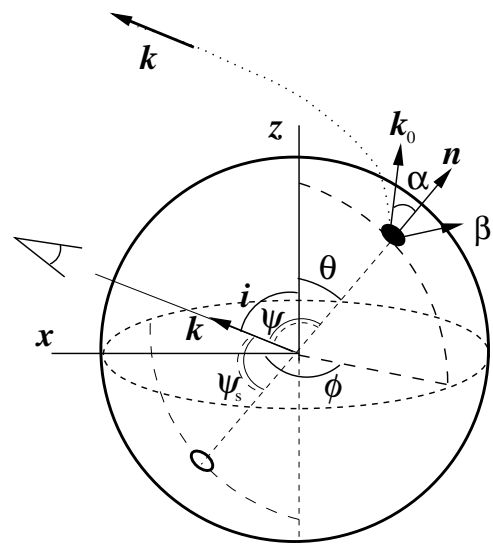,width=0.7\linewidth}}
\caption{Geometry of the problem.}
\label{fig:geom}
\end{center}
\end{figure}

\subsection{Light bending and time delay}

We choose the coordinate system so that the unit vector in the direction to the observer is $\unit{k}=(\sin i, 0, \cos i)$, where $i$ is the inclination of the spin axis to the line of sight; see Fig.~\ref{fig:geom} for the geometry. 
We set the magnetic obliquity, that is, the co-latitude of the primary hotspot, as $\theta$. 
The spot coordinates vary periodically with pulsar rotational phase $\phi$ and the unit vector of the spot normal (which is parallel to the radius-vector of the emission point in the case of a spherical star) is 
$\unit{n}=(\sin\theta \cos\phi, \sin \theta \sin\phi,\cos\theta)$. 
Thus, the angle between the direction to the spot and the line of sight is given by  
\be \label{eq:psi}
\cos\psi \equiv \unit{k}\cdot\unit{n}
= \cos i\ \cos\theta+\sin i\ \sin \theta\ \cos\phi ,
\ee
where $\phi=0$ when the spot is closest to the observer.
For the secondary spot, we substitute $\phi \rightarrow \phi+\pi$ and $\theta \rightarrow \pi - \theta$.

In the Schwarzschild metric, the photon orbits are planar and the original direction of the photon momentum near the stellar surface, $\unit{k}_0$, is related to its direction at infinity, $\unit{k}$, and the normal to the star, $\unit{n}$ \citep{PG03}:
\be\label{eq:k0}
\unit{k}_0=[ \sin\alpha \ \unit{k} + \sin(\psi-\alpha) \ \unit{n}]/\sin\psi,
\ee
where  $\alpha$ represents the angle between $\unit{k}_0$ and $\unit{n}$,  $\cos\alpha=\unit{k}_0\cdot\unit{n}$. 
The relation between angles $\alpha$ and $\psi$ can be obtained by computing an elliptical integral \citep{PFC83,B02,PG03,PB06,SNP18}: 
\be \label{eq:bend}
\psi=\int_R^{\infty} \frac{\diff r}{r^2} \left[ \frac{1}{b^2} -
\frac{1}{r^2}\left( 1- \frac{\rg}{r}\right)\right]^{-1/2} , 
\ee
where $\rg \equiv 2GM/c^2$ is the Schwarzschild radius of the NS and 
\be  \label{eq:impact}
b= \frac{R}{\sqrt{1-u}} \sin \alpha 
\ee
is the impact parameter and $u=\rg/R$ is the NS compactness. 
For our calculations here, we use approximate analytical formula for the dependence $\alpha(\psi)$ as suggested by \citet{poutanen20}: 
\be\label{eq:poutanen20app}
\cos\alpha \approx 1 - y \,(1-u)\,  G(y,u),
\ee
where $y=1-\cos\psi$ and 
\be \label{eq:Gfunc}
    G(y,u) = 1+\frac{u^2\,y^2}{112}-\frac{e}{100}u\,y\,\left[\ln\left(1-\frac{y}{2}\right)+\frac{y}{2}\right].
\ee
This approximation has a typical accuracy of  0.05\% for a large range of emission angles and for the most realistic NS compactnesses.

The spot velocity is described by the unit vector $\unit{v}=(-\sin\phi,\cos\phi,0)$ and the velocity amplitude as: 
\be\label{eq:beta}
\beta=\frac{v}{c}= \frac{2\pi R}{c} \frac{\nu}{\sqrt{1-u}} \sin\theta ,
\ee
where the observed pulsar frequency $\nu$ is corrected for the redshift $\sqrt{1-u}$.
The angle between the velocity and the photon direction is then 
\be \label{eq:cosxi}
\cos\xi = \unit{v} \cdot \unit{k}_0 = \frac{\sin\alpha}{\sin\psi}\  \unit{v} \cdot \unit{k}=  -  \frac{\sin\alpha}{\sin\psi} \sin i\ \sin\phi\  .
\ee
The unit vector of the photon momentum in the spot comoving frame can be obtained from the Lorentz transformation 
\beq 
\unit{k}^{\prime}_0&=& \Dop \left[ \unit{k}_0 - \gamma\beta \unit{v} + (\gamma-1) \unit{v} (\unit{v}\cdot \unit{k}_0 ) \right] ,
\eeq 
where  
\be \label{eq:dop}
\Dop= 1/[\gamma(1-\beta \unit{v}\cdot \unit{k}_0] = 1/[\gamma(1-\beta\cos\xi)] 
\ee
is the Doppler factor  and $\gamma =(1-\beta ^2)^{-1/2}$ is the Lorentz factor.
In the spot comoving frame, the photon momentum makes an angle $\alpha'$ with the local normal (see Appendix A in \citealt{PB06}):
\be\label{eq:aberr}
\cos\alpha'=\Dop\ \cos\alpha .
\ee
 
For rapidly rotating pulsars, we need to account for the difference in light travel time around the star.
A photon of impact parameter $b$ given by Eq. (\ref{eq:impact}) lags one with the impact parameter $b=0$ by \citep{PFC83}:
\be \label{eq:delay}
\Delta t(b)= \frac{1}{c}  \int_R^{\infty} \frac{\diff r}{1- \rg/r}
\left\{ \left[ 1-  \frac{b^2}{r^2}  \left( 1- \frac{\rg}{r} \right)
\right] ^{-1/2}  -1 \right\} .
\ee
A method to accurately compute this integral is given in \citet{SNP18}.
The corresponding phase delay is $\Delta \phi=2\pi\nu\Delta t$ and the observed phase is 
\be \label{eq:phiobs_phi}
\phi_{\rm obs}  =  \phi + \Delta \phi (\phi) .
\ee 
For analytical calculations, the phase delays for the primary and secondary spots relative to the photons emitted from the primary spot at phase $\phi=0$ can be computed approximately, assuming straight trajectories \citep{PB06}: 
\beq 
\Delta \phi_{\rm p} (\phi)  &  \! \! \approx \! \!  & 2\pi \nu R/c\   \sin i\ \sin\theta\ (1-\cos\phi), \\ 
\Delta \phi_{\rm s} (\phi) & \! \! \approx\! \!  &  2\pi \nu R/c  \left[  2\cos i\ \cos\theta + \sin i\ \sin\theta\ (1+\cos\phi) \right]. 
\eeq

\subsection{Observed flux and polarisation degree}
\label{sec:obs_flux_PD} 

The combined effect of the gravitational redshift and Doppler effect results in the following  relation between the monochromatic observed and local radiation intensities \citep[see e.g.][]{mtw73,RL79}: 
\be
I_{E} = \left (\frac{E}{E'}\right )^3 I'_{E '} (\alpha') , 
\ee
where $E/E'=\Dop \sqrt{1-u}$.
Here $I'_{E'}(\alpha')$ is the intensity at energy $E'$ and zenith angle $\alpha'$  in the frame comoving with the spot.
In order to compute the observed flux, we need to know the variations of the solid angle occupied by the spot of area $\diff S'$ in the observer's sky \citep{PG03}:
\be\label{eq:omega}
\diff\Omega=\frac{\diff S' \cos \alpha'}{D^2}\ \lensing ,
\ee
where $D$ is the source distance and 
\be\label{eq:lensing}
\lensing = \frac{1}{1-u}\  \frac{\diff\cos\alpha}{\diff\cos\psi} 
\ee
is  the lensing factor that can be computed numerically \citep{SNP18} or using analytical approximations \citep[see e.g.][]{poutanen20}. 
Hence the monochromatic flux becomes:
\be \label{eq:fluxmono}
F_{E}=I_E \diff \Omega=
(1-u)^{3/2} \Dop^{4}\  I'_{E'}(\alpha') \cos\alpha \ \lensing\  \frac{\diff S'}{D^2} ,
\ee
where we used the aberration formula (\ref{eq:aberr}).
Because the PD does not change along photon trajectory, the observed one is equal to the measured one in the spot frame at energy of the emitted photon $E'$  and at zenith angle $\alpha'$: 
\be \label{eq:polar_mono}
P_{E}= P_{E '}(\alpha') . 
\ee

\subsection{Combining Stokes parameters}
\label{sec:Stokes}

For each infinitely small spot, Eq.~(\ref{eq:fluxmono}) gives us the flux $F_E$ and Eq.~(\ref{eq:polar_mono}) provides the PD $P_E$. 
The PA $\chi$ will be derived in Sect.~\ref{sec:obs_PA}. 
The observed Stokes vector (we do not consider sources of circular polarisation) is then 
\be 
F_E \left(\begin{array}{c}  1  \\ P_E\cos2\chi \\ P_E \sin2\chi  \end{array}
\right) .
\ee 
To compute the total observed Stokes vector from two small antipodal spots, we combine the Stokes parameters from each spot accounting for different time delays. 
The PD of the total radiation is 
\be\label{eq:poltot}
P_{\rm tot} = \frac{\sqrt{(F^{\rm p}P^{\rm p})^2+(F^{\rm s}P^{\rm s})^2 + 2F^{\rm p}F^{\rm s}P^{\rm p}P^{\rm s}
\cos (2\chi^{\rm p}- 2\chi^{\rm s})}}{F^{\rm p}+F^{\rm s}},
\ee
where superscripts, p and s, denote the primary and secondary spot, respectively, and we skipped the energy subscript $E$ for clarity. 
The total PA is given by
\be \label{eq:patot}
\tan (2\chi_{\rm tot}) = \frac{F^{\rm p}P^{\rm p} \sin 2\chi^{\rm p} +
F^{\rm s}P^{\rm s}  \sin 2\chi^{\rm s}}
{F^{\rm p}P^{\rm p} \cos 2\chi^{\rm p} + F^{\rm s}P^{\rm s} \cos 2\chi^{\rm s}} .
\ee

If the spots are extended, we can integrate over the spot surface using angular coordinates in the comoving frame \citep{NP18,Lo18,Bogdanov19L26} 
\be 
\diff S'=\gamma\ R^2\ \diff \cos\theta\ \diff\phi' . 
\ee 
The surface element will be defined by the latitude and the azimuthal angle. 
For each element, we compute the dependence of the Stokes parameters on the observed phase, accounting  for the difference in latitude, time delay, and PA, and sum them up. 

\section{Polarisation angle}
\label{sec:obs_PA} 

\subsection{Rotating vector model}

In order to describe polarisation observed from an infinitely small spot, we introduce the main polarisation basis,
\beq  \label{eq:basis_main1}
\unit{e}_1^{\rm m}&=&\frac{\unit{\omega} - \cos i\ \unit{k} }{\sin i} = (-\cos i,0,\sin i ), \\
 \label{eq:basis_main2}
\unit{e}_2^{\rm m}&=&\frac{\unit{k} \times \unit{\omega}}{\sin i} = (0,-1,0),
\eeq
where $\unit{\omega}=(0,0,1)$ denotes the unit vector along the stellar rotational axis.

In the absence of a relativistic rotation of the polarisation plane (for a slowly rotating star), the polarisation vector lies in the plane formed by the local normal to the spot, $\unit{n}$, and the direction to the observer, $\unit{k}$.
The corresponding polarisation basis is
\be \label{eq:basis_slow}
\unit{e}_1=\frac{\unit{n} - \cos \psi\ \unit{k} }{\sin \psi}, \quad
\unit{e}_2=\frac{\unit{k} \times \unit{n} }{\sin \psi} .
\ee
The PA $\chi_0$ measured from the projection of the spin axis on the plane of the sky in the counter-clockwise direction is given by:
\beq\label{eq:coschi0}
\cos \chi_0 \!\! & \!\!= \!\!&\!\! \unit{e}_1^{\rm m} \cdot \unit{e}_1 = \unit{e}_2^{\rm m} \cdot \unit{e}_2
 =  \frac{\sin i\ \cos \theta - \cos i\  \sin \theta\  \cos \phi }
{\sin \psi},  \\
\sin \chi_0 \!\! &\!\! =\!\! &\!\! \unit{e}_2^{\rm m} \cdot \unit{e}_1 = - \unit{e}_1^{\rm m} \cdot \unit{e}_2
= - \frac{\sin \theta\ \sin \phi} {\sin \psi} .
\eeq
We thus get the expression for the PA as in the rotating vector model (RVM) of \citet{RC69}: 
\be \label{eq:pa_rvm}
\tan\chi_0= \frac{\sin \theta\ \sin \phi}
{-\sin i\ \cos \theta  + \cos i\ \sin \theta\  \cos \phi }.
\ee

\subsection{Rotation of PA due to rapid rotation}
 
\subsubsection{The PA for rapidly rotating NS when ignoring bending}

Now we evaluate the effect of rapid rotation  while neglecting light bending, that is,  accounting for the SR effects only. 
The photon momentum direction in the frame corotating with the spot is
 \beq 
\unit{k'}&=& \Dop \left[ \unit{k} - \gamma\beta \unit{v} + (\gamma-1) \unit{v} (\unit{v}\cdot \unit{k} ) \right] \nonumber \\
&=& \Dop \left( \begin{array}{c} \sin i + \sin\phi \ [\gamma\beta-(\gamma-1)\cos\xi] \\
- \cos\phi \ [\gamma\beta-(\gamma-1)\cos\xi] \\
 \cos i \\ \end{array}  \right) ,
\eeq 
where now  $\cos\xi=-\sin i \sin\phi$, that follows from Eq.\,\eqref{eq:cosxi}, with $\alpha=\psi$, because light bending is ignored. 
Following  \citet{Fer73,Fer76} we introduce a common vector which does not change upon Lorentz transformation from the comoving frame related to the spot to the laboratory frame related to the observer. 
This vector is normal to both the photon momentum and the vector of spot velocity: 
\beq \label{eq:comm_vector}
\unit{M} & =&  \frac{\unit{k} \times \unit{v}  }{|\unit{k} \times \unit{v} |} =  \frac{\unit{k'} \times \unit{v}  }{|\unit{k'}\times \unit{v} |} \nonumber \\
& =& {\left( -\cos i \cos \phi ,  -\cos i \sin \phi, \sin i \cos\phi \right) } / { \sin \xi }.   
\eeq
The polarisation basis related to the common vector is then 
\be
\unit{e}^{\rm M}_1=\unit{M}, \quad
\unit{e}^{\rm M}_2= \unit{k} \times \unit{M}  . 
\ee
The position angle of the common vector as measured in the main basis is 
\beq
\cos\chi_{\rm M} & = &\unit{e}^{\rm m}_1 \cdot \unit{e}^{\rm M}_1 = \frac{\cos\phi}{\sin\xi} , \\
\sin\chi_{\rm M}  & = & \unit{e}^{\rm m}_2 \cdot \unit{e}^{\rm M}_1 = \frac{\cos  i\ \sin \phi}{\sin\xi} , 
\eeq
and then 
\be \label{eq:tanchiM}
\tan \chi_{\rm M} =\cos i\ \tan \phi.
\ee  
For slow rotation, the polarisation plane is defined by the vector $\unit{e}_1$ and the angle $\Delta$ that it makes with the common vector basis vector $\unit{e}^{\rm M}_1$ and is given by
\beq
\cos\Delta & = &\unit{e}^{\rm M}_1 \cdot \unit{e}_1 = \frac{-\cos i \sin \theta+ \sin i \cos \theta \cos\phi}{\sin\xi \sin\psi} , \\
\sin\Delta & = &- \unit{e}^{\rm M}_1 \cdot \unit{e}_2 
= \frac{\cos \xi \cos\psi}{\sin\xi \sin\psi} 
= - \frac{\sin  i \sin \phi \cos\psi}{\sin\xi \sin\psi} , 
\eeq
and 
\be \label{eq:tanDelta}
\tan \Delta = \frac{\sin  i \sin \phi \cos\psi}{\cos i \sin \theta- \sin i \cos \theta \cos\phi} . 
\ee  
It is easy to check that the PA $\chi= \chi_{\rm M} + \Delta= \chi_0$, as given by Eq.~(\ref{eq:pa_rvm}). 

In the case of rapid rotation, the angle that the polarisation plane makes with the common vector should be measured in the spot comoving frame. 
We introduce the polarisation basis in that frame: 
\be
\unit{e}'_1=\frac{\unit{n} - \cos \psi'\ \unit{k'} }{\sin \psi'}, \quad
\unit{e}'_2=\frac{\unit{k'}\times \unit{n} }{\sin \psi'} ,
\ee
where $\cos \psi' = \unit{n} \cdot \unit{k'} = \Dop \cos\psi$. 
Then the angle $\Delta'$ made by the polarisation vector with the common vector as seen in the comoving frame is
\beq
\cos\Delta' & = &\unit{e}^{\rm M}_1 \cdot \unit{e}'_1 = \frac{-\cos i \sin \theta+\sin i \cos \theta \cos\phi}{\sin\xi \sqrt{1-\Dop^2\cos^2\psi}} , \\
\sin\Delta' & = &- \unit{e}^{\rm M}_1 \cdot \unit{e}'_2 
= \frac{\cos \xi' \cos\psi}{\sin\xi \sin\psi'} \nonumber \\
&= & \frac{\cos\psi}{\sin\xi \sqrt{1-\Dop^2\cos^2\psi}} \frac{\cos\xi-\beta}{1-\beta\cos\xi} , 
\eeq 
and 
\be \label{eq:tanDeltapr}
\tan \Delta' = \frac{\cos\psi}{\sin i \cos \theta \cos\phi-\cos i \sin \theta}  \frac{\cos\xi-\beta}{1-\beta\cos\xi} . 
\ee  
Thus, the observed PA is the sum $\chi= \chi_{\rm M} + \Delta'$ and using Eqs.~(\ref{eq:tanchiM}) and (\ref{eq:tanDeltapr}), we get 
\be \label{eq:tanchi_rel}
\tan \chi = \frac{ \sin \theta  \sin \phi + \beta\, (\sin i \sin \theta+\cos i \cos \theta \cos\phi) } 
{-\sin i \cos \theta + \cos i \sin \theta \cos\phi  - \beta\cos \theta \sin\phi}   . 
\ee    
The deviation of the PA from that given by the non-relativistic RVM  (\ref{eq:pa_rvm}) and owing to the SR effects only is then simply: 
 \be \label{eq:tanchic}
\tan \chi_{\rm c, SR}  \! =\!  \tan ( \chi\!-\!\chi_0) \!    =\!  \beta \cos\psi 
\frac{\cos i\, \sin \theta - \sin i \cos \theta \cos \phi} {\sin^2 \psi + \beta \sin i\, \sin\phi}   . 
\ee    

We note that this expression can also be used to get the rotation of the polarisation plane of the accretion disc radiation. 
Substituting $\theta=0$ and $\psi\rightarrow i$, we get 
 \be \label{eq:tanchic_disc}
\tan \chi_{\rm disc}  =
-  \frac{\beta\ \cos i\ \cos \phi} {\sin i + \beta  \sin\phi}   ,
\ee    
which is equivalent to Eq.\,(18) from \citet{Connors80}. 

\subsubsection{The PA for rapidly rotating NS accounting for bending in the Schwarzschild metric}
\label{sec:PA_fast_bend} 

Because of the light bending, the photon direction close to the NS surface, $\unit{k}_0$, differs from that at infinity, $\unit{k}$. 
Therefore, the common vector should be defined by $\unit{k}_0$ and  the spot velocity vector:  
\beq \label{eq:comm_vector_rel}
\unit{M} & =&  \frac{\unit{k}_0 \times \unit{v}  }{|\unit{k}_0 \times \unit{v} |}  
= \frac{\unit{k}^{\prime}_0 \times \unit{v}  }{|\unit{k}^{\prime}_0\times \unit{v} |}  \\
& =&\frac{1}{\sin\xi \sin\psi} \left( \begin{array}{c} 
-\cos \phi\ [\cos i \sin \alpha + \cos\theta \sin(\psi-\alpha)] \\
-\sin \phi\ [\cos i \sin \alpha + \cos\theta \sin(\psi-\alpha)] \\
 \sin i \cos \phi  \sin \alpha + \sin\theta \sin(\psi-\alpha)  
\end{array}  \right) ,      \nonumber
\eeq
where now (see Eq.\,\eqref{eq:cosxi}) 
\be \label{eq:sinxi}
\sin\xi =  \sqrt{ 1 -  \frac{\sin^2\alpha}{\sin^2\psi} \sin^2 i\ \sin^2\phi\ } .
\ee
Here, we introduce the polarisation basis related to photon momentum in the comoving frame, $\unit{k}^{\prime}_0$, and the surface normal:
\be
\unit{e}_1^{\prime\,0}=\frac{\unit{n} - \cos \alpha'\ \unit{k}^{\prime}_0 }{\sin \alpha'}, \quad
\unit{e}_2^{\prime\,0}=\frac{\unit{k}^{\prime}_0 \times \unit{n} }{\sin \alpha'}  .
\ee
The position angle of  the common vector $\unit{M}$ as viewed in the comoving frame  is 
\beq
\cos\chi'_{\rm M} & = & \unit{M}  \cdot \unit{e}_1^{\prime\,0}= 
\frac{\sin \alpha}{\sin\psi} \frac{\sin i\cos\theta\cos \phi-\cos i\sin\theta}{\sin\xi\sqrt{1-\Dop^2\cos^2\alpha}} , 
\\
\sin\chi'_{\rm M} & = & 
- \unit{M}  \cdot \unit{e}_2^{\prime\,0}
= \frac{\cos \xi' \cos\alpha}{\sin\xi \sin\alpha'} \nonumber \\
&= & \frac{\cos\alpha}{\sin\xi \sqrt{1-\Dop^2\cos^2\alpha}} \frac{\cos\xi-\beta}{1-\beta\cos\xi} , 
\eeq 
and, correspondingly, 
\be \label{eq:tan_chiprM} 
\tan \chi'_{\rm M} = \frac{\sin\psi\cos\alpha }{\sin\alpha(\sin i \cos \theta \cos\phi-\cos i \sin \theta)}  \frac{\cos\xi-\beta}{1-\beta\cos\xi} . 
\ee

\begin{figure}
\begin{center}
\centerline{\epsfig{file=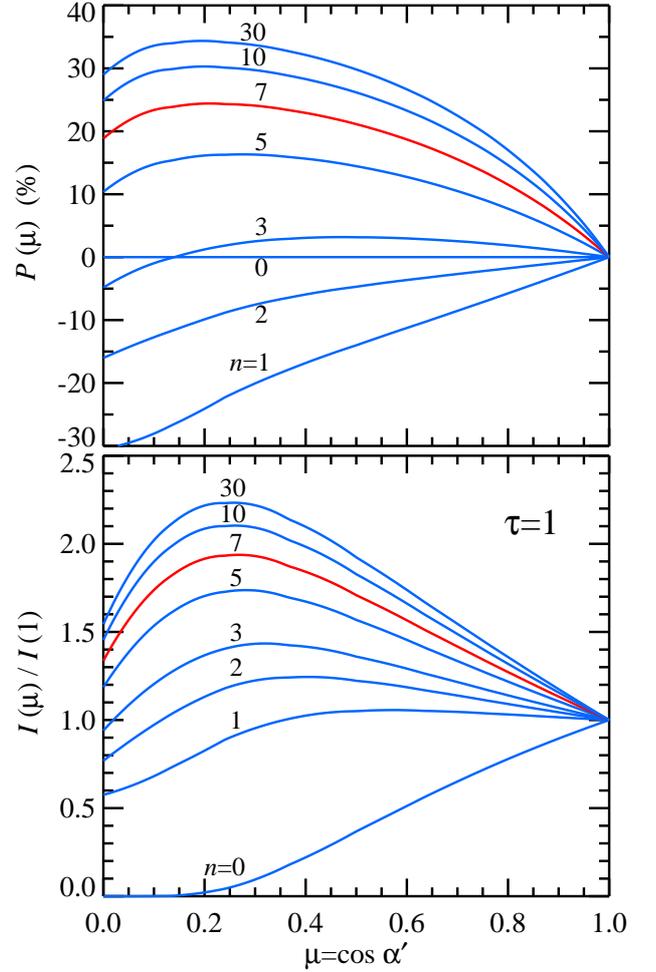,width=0.90\linewidth}}
\caption{Angular distribution of intensity (\textit{lower panel}) and PD (\textit{upper panel}) for different scattering orders for a slab of Thomson optical depth $\tau=1$.}
\label{fig:Thomson}
\end{center}
\end{figure}

Now we can evaluate the PA $\chi_{\rm M}$ of vector $\unit{M}$ from the projection of the stellar normal on the sky as viewed in the laboratory frame. 
The corresponding polarisation basis is  
\be
\unit{e}_1^0=\frac{\unit{n} - \cos \alpha\ \unit{k}_0 }{\sin \alpha},
 \quad
\unit{e}_2^0=\frac{\unit{k}_0 \times \unit{n} }{\sin \alpha} =  \unit{e}_2.
\ee
The equality $\unit{e}_2^0=\unit{e}_2$ (defined by Eq.\,(\ref{eq:basis_slow})) is related to the fact that in Schwarzschild metric photon trajectories are planar.
We get 
\beq
\cos\chi_{\rm M} & = & 
\unit{e}_1^{0}  \cdot \unit{M} 
=   \frac{\sin i\cos\theta\cos \phi-\cos i\sin\theta}{\sin\xi\sin\psi} , 
\\
\sin\chi_{\rm M} & = & 
\unit{e}_2^{0}   \cdot \unit{M} 
=  - \frac{\cos \xi \cos\alpha}{\sin\xi \sin\alpha} 
=  \frac{\sin i  \sin\phi \cos\alpha }{\sin\xi \sin\psi}  , 
\eeq 
and 
\beq \label{eq:tan_chiM_rel} 
\tan \chi_{\rm M} &=& - \frac{\sin\psi\cos\alpha }{\sin\alpha(\sin i \cos \theta \cos\phi-\cos i \sin \theta)}\cos\xi  \nonumber\\ 
&=& \frac{\sin i\, \sin\phi \cos\alpha }{\sin i \cos \theta \cos\phi-\cos i \sin \theta} . 
\eeq  

\begin{figure}
\begin{center}
\centerline{\epsfig{file=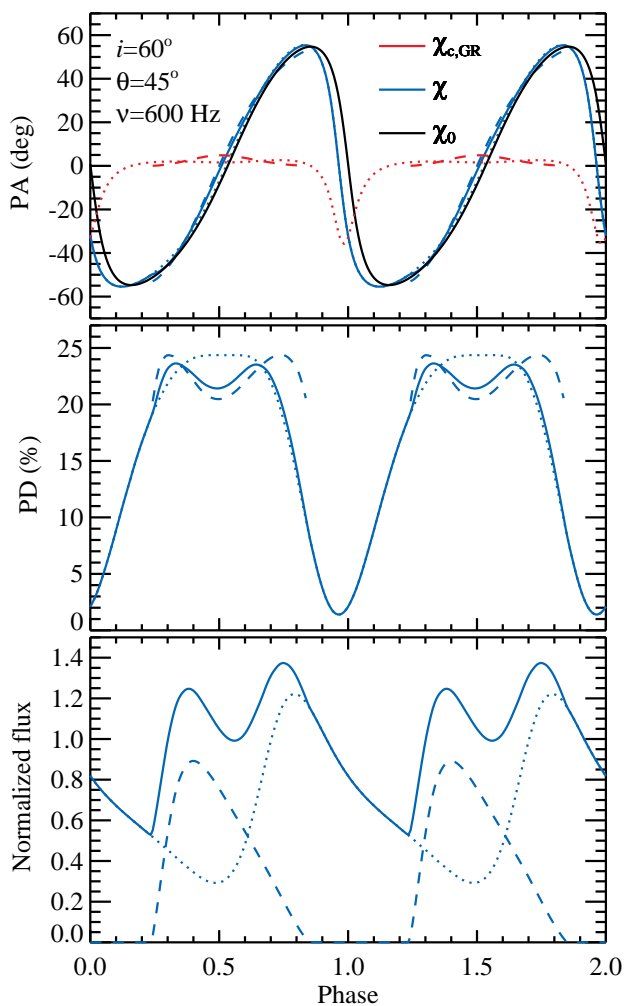,width=0.90\linewidth}}
\caption{Normalised flux (\textit{lower panel}), PD (\textit{middle panel}), and PA (\textit{upper panel}) as a function of pulsar phase for the fiducial set of parameters $M=1.5\msun$, $R=12$~km, $\nu=600$~Hz, the inclination $i=60\degr,$ and the magnetic obliquity is $\theta=45\degr$. 
The dotted lines show the contribution from the primary spot, the dashed lines correspond to the secondary spot, and the solid lines indicate the results for both spots. 
The PA $\chi_0$ from the RVM, as given by Eq.\,(\ref{eq:pa_rvm}) for the primary spot, is shown by the black line. 
The red lines indicate the GR correction to the PA $\chi_{\rm c,GR}$ given by Eq.\,(\ref{eq:tan_chic_rel}) for the two spots. 
The blue lines show the total PA $\chi=\chi_0+\chi_{\rm c,GR}$ for both spots separately and together. 
}
\label{fig:fiducial}
\end{center}
\end{figure}

Thus, because of the spot motion and light bending, the GR correction to the PA, $\chi_{\rm c,GR} = \chi'_{\rm M} + \chi_{\rm M}$, is 
\be\label{eq:tan_chic_rel}
\tan \chi_{\rm c,GR}  =    \beta \cos\alpha \frac{\sin \alpha}{\sin\psi }  
\frac{\cos i\ \sin \theta - \sin i\ \cos \theta\ \cos \phi} {\sin^2\alpha\ - \beta \cos\xi} .
\ee 
This coincides with expression (29) of \citet{VP04}.
The observed PA of the polarisation plane relative to the projection of the stellar spin on the sky is then $\chi=\chi_0+\chi_{\rm c,GR}$: 
\be \label{eq:tanchi_rel_bend}
\tan \chi = 
\frac{ \sin \theta  \sin \phi + \beta \ A  } 
{-\sin i \cos \theta + \cos i \sin \theta \cos\phi  -  \beta\sin\phi \ C }, 
\ee    
where 
\beq  \label{eq:aux_rel_bend}
A&=& \frac{\sin\psi}{\sin\alpha} B+  \frac{\cos\alpha-\cos\psi}{\sin\alpha \sin\psi} (\cos\phi-B\cos\psi) , \nonumber \\ 
B&=& \sin i \sin \theta+\cos i \cos \theta \cos\phi, \\
C&=& \frac{\sin\psi}{\sin\alpha}\cos \theta +  \frac{\cos\alpha-\cos\psi}{\sin\alpha \sin\psi} (\cos i -\cos \theta \cos\psi) .  \nonumber
\eeq
For slow rotation, $\beta\rightarrow 0$, Eq.\,(\ref{eq:tanchi_rel_bend}) transforms to Eq.\,(\ref{eq:pa_rvm}), while if we ignore light bending, that is, by placing $\alpha=\psi$ in Eqs.\,(\ref{eq:aux_rel_bend}), the equation transforms to Eq.\,(\ref{eq:tanchi_rel}).

\begin{figure}
\begin{center}
\centerline{\epsfig{file=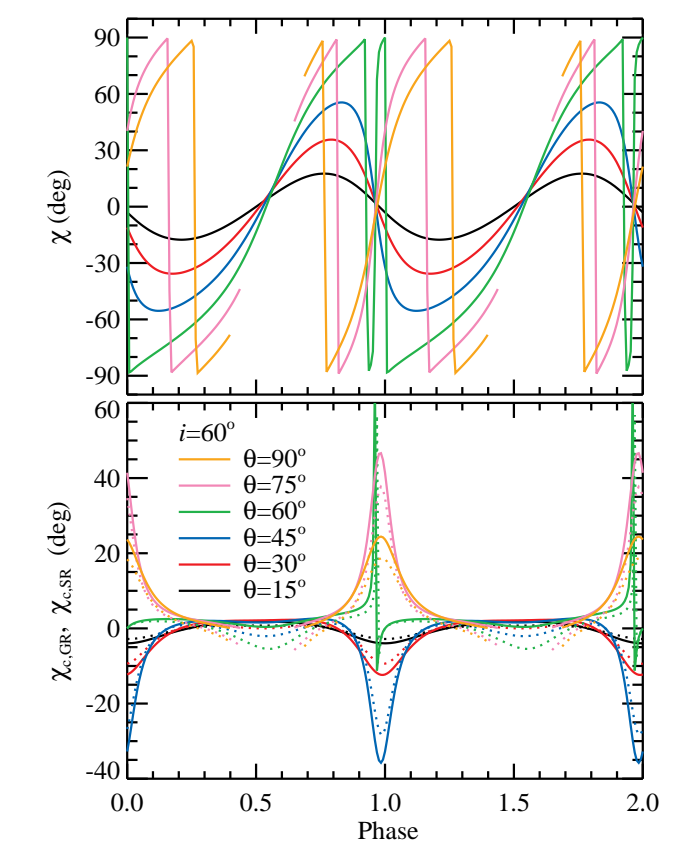,width=0.90\linewidth}}
\caption{\textit{Upper panel}: Relativistic PA as a function of pulsar phase for one small spot at different co-latitudes $\theta$=15\degr, 30\degr, 45\degr, 60\degr, 75\degr, and 90\degr\  is shown with black, red, blue, green, pink, and orange lines, respectively. 
NS parameters are $M=1.5\msun$, $R=12$~km, $\nu=600$~Hz and inclination $i=60\degr$. 
The spot is not visible at some phases for $\theta$=75\degr and 90\degr.
\textit{Lower panel}: Relativistic corrections to the PA. 
The dotted lines indicate the SR correction angle $\chi_{\rm c,SR}$, which  accounts for special relativity only. 
The solid lines correspond to the GR correction $\chi_{\rm c,GR}$, which accounts for both light bending and SR effects.
}
\label{fig:chi_theta}
\end{center}
\end{figure}

\section{Application to accreting ms pulsars}

As an example of the application of the developed formalism, we consider one or two antipodal small spots which emit polarised radiation. 
We assume that there is a plane-parallel electron-scattering dominated atmosphere of Thomson optical depth $\tau=1$ atop the cold NS surface. 
This setup may be associated with the flat accretion shock above the NS surface. 
Incident photons of energy $E_0$ from the bottom of the slab have isotropic intensity. 
They are scattered multiple times in the slab. 
The angular distribution of the emergent radiation and its polarisation characteristics are described in \citet{VP04}. 
In a realistic situation, the electrons in the slab are hot (according to observations $kT_{\rm e}\approx$30--70~keV, see e.g. review in \citealt{P06}) and up-scatter  incident photons to higher energies. 
We can relate each scattering order $n$ to the photon energy $E'$ by the following relation \citep{RL79}
\be 
E'=E_0 \left(1 + 4\frac{kT_{\rm e}}{m_{\rm e} c^2} \right)^n .
\ee
The emergent intensity and PD for different scattering orders is shown in Fig.\,\ref{fig:Thomson}. 
The PD of radiation emergent from the slab is positive if the dominant electric vector oscillations lie in the meridional plane defined by the spot normal and the direction of photon propagation and it is negative if the oscillations are perpendicular to that plane. 

Because the main emphasis of this paper is the variation of the PA with the phase, along with the impact of rapid rotation on PA, we simply choose  $n=7$ for the purposes of illustration (see red lines in Fig.\,\ref{fig:Thomson}). 
This choice does not affect the PA.

\begin{figure}
\begin{center}
\centerline{\epsfig{file=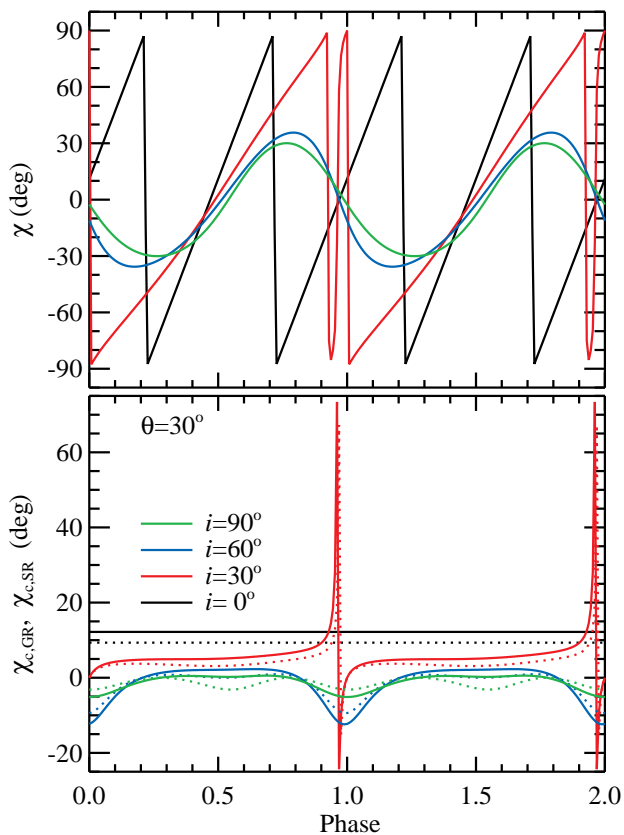,width=0.90\linewidth}}
\caption{Same as Fig.\,\ref{fig:chi_theta}, but for magnetic obliquity  $\theta =30\degr$ and different inclinations $i$=0\degr, 30\degr, 60\degr, and 90\degr, shown by the black, red, blue, and green lines, respectively. 
}
\label{fig:chi_incl}
\end{center}
\end{figure}

We select the fiducial set of parameters: $M=1.5\msun$, $R=12$~km, $\nu=600$~Hz, inclination $i=60\degr$, and magnetic obliquity $\theta=45\degr$.  
We consider infinitely small spots in most of the calculations.  
The dependence of the flux (normalised to unity), PD, and PA on the observed pulsar phase for the fiducial set are shown in Fig.\,\ref{fig:fiducial}. 
For the considered geometry and NS compactness, the primary spot is visible at all phases, while the secondary spot is seen only during certain intervals. 
The pulse profile has a double-peak structure. 
The PD has a minimum when the viewing angle to the spot $\alpha'$ is close to 0. 
Because the flux maximum is reached at a larger angle with $\mu=\cos\alpha'\approx0.2$, the phase of the maximum flux differs from the phase of the minimum PD. 
The PA follows an S-like curve.  
We see that any additional rotation of the PA introduced by relativistic motion and light bending (red lines) reaches, in this case, 40\degr. 
The effect is largest for the primary spot, which passes close to the line connecting the observer to the NS centre. 
In the following examples, we vary individual parameters to view the effect on the PA.  

Figure\,\ref{fig:chi_theta} shows the PA for a single spot at different magnetic obliquities $\theta$ and fixed observer inclination. 
We see that for a small $\theta$, when the spot velocity $\beta$ is low, the correction to the PA owing to the GR effects $\chi_{\rm c,GR}$ is small (see bottom panel) and the total PA (top panel) is close to a sine-wave. 
Increasing $\theta$ leads to a higher velocity and larger $\chi_{\rm c,GR}$. 
The SR correction to the PA (dotted curves) has a very similar structure, but lower amplitude. 
The corrections are highest when $i=\theta$ and the spot passes through the line of sight (green curves in the bottom panel). 
We see from Eqs.\,(\ref{eq:tanchic}) and  (\ref{eq:tan_chic_rel}) that the tangent of the correction angle becomes infinite (i.e. correction angle is $\pm \pi/2$) when the denominator is zero. 
For the SR correction, this happens at an emission phase $\phi\approx -\beta/\sin i$. 
The denominator in  Eq.\,(\ref{eq:tan_chic_rel}) is zero at a slightly different phase because of light bending. 
Approximating  $\sin\alpha/\sin\psi\approx\sqrt{1-u}$ \citep{PB06}, we get $\phi\approx -\beta/(\sin i\sqrt{1-u})$. 
The observed phase in this case is nearly the same because the time delays here are computed relative to trajectories with zero impact parameter.

\begin{figure}
\begin{center}
\centerline{\epsfig{file=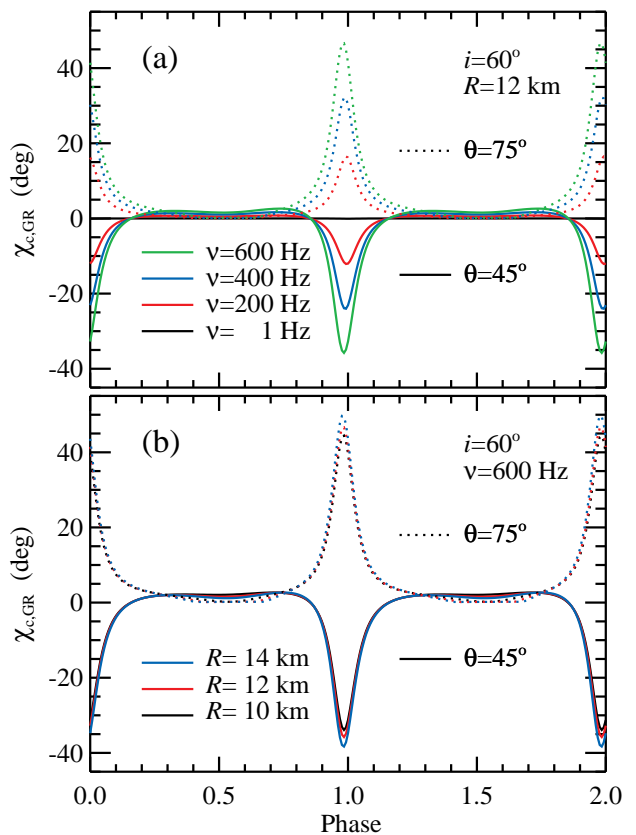,width=0.90\linewidth}}
\caption{\textit{Panel a}: Phase dependence of the PA GR corrections for different NS spin rates. 
Black, red, blue, and green lines show the results for $\nu=$1, 200, 400 and 600~Hz. 
The solid and dotted lines correspond to magnetic obliquity $\theta=45\degr$ and 75\degr, respectively. 
Other parameters are $M=1.5\msun$, $R=12$~km, and inclination $i=60\degr$.
Solid and dotted lines correspond to $\theta =45\degr$ and 75\degr, respectively. 
\textit{Panel b}: Same as \textit{panel a}, but for different NS radii. Black, red, and blue lines show the results for $R=$10, 12, and 14~km for spin $\nu=$600~Hz.
}
\label{fig:chi_spin_rad}
\end{center}
\end{figure}

The phase dependencies of the PA for different inclinations and fixed $\theta=30\degr$ are shown in Fig.\,\ref{fig:chi_incl}. 
We see again that the relativistic correction (bottom panel) can be very large when $i=\theta$ (red lines). 
Interestingly, even for zero inclination (black lines), the PA does not follow the RVM dependence $\tan\chi_0=\tan\phi$, but there is a shift in the PA determined by simple relations
$\chi_{\rm c,SR}\approx \beta \cot\theta$ and $\chi_{\rm c,GR}\approx\beta \cot\alpha$ (where $\alpha<\theta$ is determined by the light bending formula with $\psi=\theta$, see Eqs.\,\ref{eq:bend}--\ref{eq:poutanen20app}). 

Figure\,\ref{fig:chi_spin_rad}a shows the phase dependence of the GR correction to the PA for different NS spins. 
Obviously, for a slowly rotating star with $\nu=1$~Hz, the correction is negligibly small. 
It grows nearly linearly with the rotation rate as follows from Eq.\,\eqref{eq:tan_chic_rel}.
The sign of the correction depends on the viewing geometry.
At phase $\phi\approx 0$, when the correction is large, it scales as  $\sin(\theta-i)$. 
This results in a predominantly positive $\chi_{\rm c,GR}$ for $\theta>i$ and predominantly negative  for $\theta<i$. 

The dependence of the GR correction to the PA on a NS radius for a rapidly spinning star is illustrated in Fig.\,\ref{fig:chi_spin_rad}b.
At a higher NS radius, the velocity of the spot is higher, resulting in a larger correction, because it scales nearly linearly with $\beta\propto R$. 
The PA correction depends also on the NS compactness, which influences the light bending. 
However, we see that the effect of the radius on $\chi_{\rm c,GR}$ is rather small.

\begin{figure}
\begin{center}
\centerline{\epsfig{file=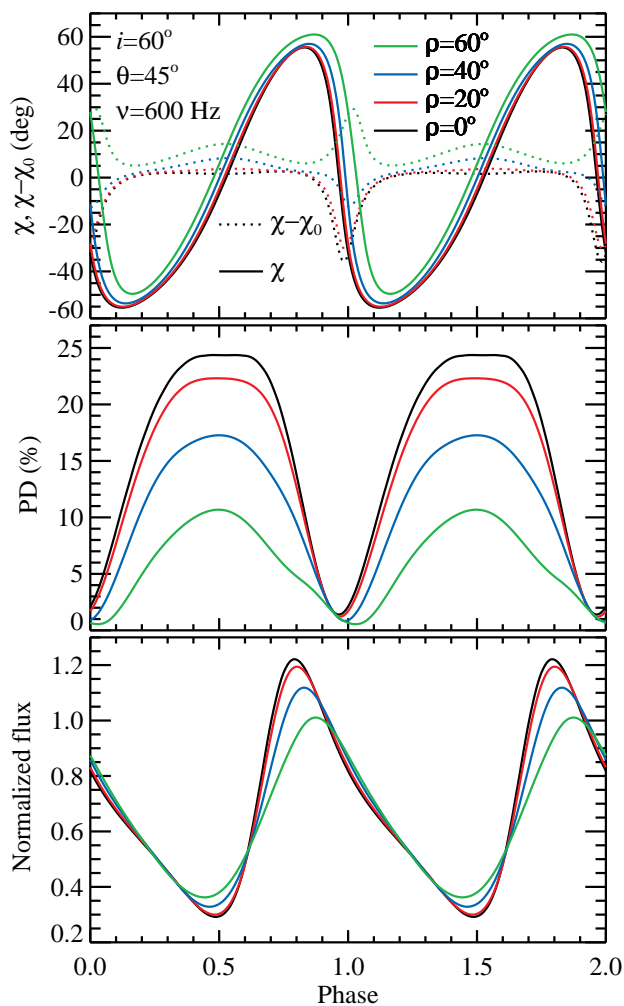,width=0.90\linewidth}}
\caption{Normalised flux (\textit{lower panel}), PD (\textit{middle panel}), and PA (\textit{upper panel}) as a function of pulsar phase for the fiducial set of parameters for one spot of different sizes.  
The black, red, blue, and green lines correspond to the spot angular radius $\rho$=0\degr, 20\degr, 40\degr, and 60\degr. 
The solid lines at the upper panel show the relativistic PA $\chi$, while the dotted lines indicate its difference from the RVM value $\chi_0$. 
}
\label{fig:all_rho}
\end{center}
\end{figure}

All above results are obtained for an infinitely small spot. 
Here, the question arises whether the PA would be affected so much by relativistic effects if the spot is large. 
We have computed the observed flux, PD, and PA for the fiducial set of parameters and varying the size of the spot from zero to 60\degr.  
We note here that calculations of the Stokes vector from an extended spot involves integration over the spot surface, accounting for 
variations in latitude, azimuthal angle, vector of the local normal, polarisation bases, and time delays; see Sect.\,\ref{sec:Stokes}. 
The results are shown in Fig.\,\ref{fig:all_rho}.
We see that with increasing spot size, the harmonic content of the flux gets reduced. 
Also the PD is reduced for a larger spot by a factor of two because different spot elements are observed at different angles.  
Interestingly, the PA is affected much less. 
We see a large deviation of the relativistic PA $\chi$ from the values predicted by the RVM $\chi_0$. 
For the chosen set of parameters, deviations decrease up to the spot angular radius $\rho\approx 50\degr$ and then they change sign and start increasing again in absolute value.

\section{Summary}

In this paper, we develop a relativistic rotating vector model. 
We derive exact analytic expressions for the PA observed from a spot on a rapidly rotating spherical NS using the so called common vector formalism. 
The expressions for PA are given when only SR effects are accounted for. 
We also derive formulae for PA accounting for the light bending in the Schwarzschild metric. 

We compute relativistic PA for different inclinations and magnetic obliquities, various spin rates, and NS compactnesses. 
We show that deviation of relativistic PA from the standard RVM may be very large, especially when the spot passes close to the line-of-sight towards the NS centre. 
These deviations grow nearly linearly with the spin rate but are less sensitive to the NS radius. 
Even for large spots, the deviations are significant and may reach tens of degrees. 

The developed formalism can be applied to compute waveforms and polarisation profiles from millisecond pulsars.  
Observations in the X-ray range with the upcoming polarimetric missions such as {IXPE} and {eXTP} will serve as a powerful tool for determining the geometry of the emission region in rapidly rotating NSs showing coherent millisecond oscillations.
Combined with the waveform analysis, they may improve  constraints on the NS mass and radius and the equation of state of cold dense matter.

\begin{acknowledgements}
This research was supported by the Russian Science Foundation grant 20-12-00364, the Academy of Finland grants 322779 and 333112, and the Magnus Ehrnrooth foundation. 
I thank Valery Suleimanov and Tuomo Salmi for valuable comments. 
\end{acknowledgements}


\end{document}